\documentclass[twocolumn,tighten,times]{nature}
\usepackage{graphicx,epsfig,amssymb,amsmath}
\usepackage{epstopdf}
\usepackage{caption}

\usepackage[T1]{fontenc}
\usepackage{threeparttable}

\usepackage[colorlinks=true,breaklinks=true,linkcolor=magenta,citecolor=blue,urlcolor=green]{hyperref}

\newcommand\figpsr{3}

\newcommand\cts{counts~s$^{-1}$}

\newcommand\apj{{Astrophys.\ J.}}%
\newcommand\apjl{{Astrophys.\ J.}}%
\newcommand\aap{{Astron.\ Astrophys.}}%
\newcommand\mnras{{Mon.\ Not.\ R.\ Astron.\ Soc.}}%
\newcommand\nat{{Nature}}%
\newcommand\pasp{{Publ.\ Astron.\ Soc.\ Pac.}}%

\title{Re-detection and a Possible Time Variation of Soft X-ray Polarisation from the Crab}

\author{Hua Feng$^{1,2\ast}$,
Hong Li$^{2}$,
Xiangyun Long$^{2}$,
Ronaldo Bellazzini$^{3}$,
Enrico Costa$^{4}$,
Qiong Wu$^{2}$,
Jiahui Huang$^{2}$,
Weichun Jiang$^{5}$,
Massimo Minuti$^{3}$,
Weihua Wang$^{6}$,
Renxin Xu$^{6}$,
Dongxin Yang$^{2}$,
Luca Baldini$^{3}$,
Saverio Citraro$^{3}$,
Hikmat Nasimi$^{3}$,
Paolo Soffitta$^{4}$,
Fabio Muleri$^{4}$,
Aera Jung$^{2}$,
Jiandong Yu$^{7}$,
Ge Jin$^{8}$,
Ming Zeng$^{2}$,
Peng An$^{7}$,
Alessandro Brez$^{3}$,
Luca Latronico$^{9}$,
Carmelo Sgro$^{3}$,
Gloria Spandre$^{3}$ \&
Michele Pinchera$^{3}$}

\begin{document}

\maketitle

\begin{affiliations}
\item Department of Astronomy, Tsinghua University, Beijing 100084, China
\item Department of Engineering Physics, Tsinghua University, Beijing 100084, China
\item INFN-Pisa, Largo B. Pontecorvo 3, 56127 Pisa, Italy
\item IAPS/INAF, Via Fosso del Cavaliere 100, 00133 Rome, Italy
\item Key Laboratory for Particle Astrophysics, Institute of High Energy Physics, Chinese Academy of Sciences, Beijing 100049 
\item Department of Astronomy, School of Physics, Peking University, Beijing 100871, China
\item School of Electronic and Information Engineering,  Ningbo University of Technology, Ningbo, Zhejiang 315211, China
\item North Night Vision Technology Co., Ltd., Nanjing 211106, China
\item INFN, Sezione di Torino, Via Pietro Giuria 1, I-10125 Torino, Italy
\item [$^\ast$] email: hfeng@tsinghua.edu.cn
\end{affiliations}

% https://www.nature.com/nature/for-authors/formatting-guide

%\linenumbers

\begin{abstract}

The Crab nebula is so far the only celestial object with a statistically significant detection in soft x-ray polarimetry\cite{Novick1972,Weisskopf1976,Weisskopf1978a,Silver1978}, a window that has not been explored in astronomy since the 1970s. However, soft x-ray polarimetry is expected to be a sensitive probe of magnetic fields in high energy astrophysical objects including rotation-powered pulsars\cite{Dyks2004a,Takata2007,Harding2017} and pulsar wind nebulae\cite{Bucciantini2017}. Here we report the re-detection of soft x-ray polarisation after 40 years from the Crab nebula and pulsar with PolarLight\cite{Feng2019}, a miniature polarimeter utilising a novel technique\cite{Costa2001,Bellazzini2003b} onboard a CubeSat.  The polarisation fraction of the Crab in the on-pulse phases was observed to decrease after a glitch of the Crab pulsar on July 23, 2019, while that of the pure nebular emission remained constant within uncertainty.  The phenomenon may have lasted about 100 days. If the association between the glitch and polarisation change can be confirmed with future observations, it will place strong constraints on the physical mechanism of the high energy emission\cite{Cheng1986a,Muslimov2004,Kalapotharakos2012} and glitch\cite{Baym1969,Anderson1975,Alpar1993} of pulsars. 

\end{abstract}

In the energy band of a few keV (hereafter referred to as the soft x-ray band), where the emission of most high energy astrophysical objects peaks, x-ray polarimetry is argued to be a powerful and sometimes unique tool in diagnosing the magnetic field, geometry, and emission mechanism in astrophysics\cite{Kallman2004,Soffitta2013}. However, due to technical difficulties, soft x-ray polarimetry has been a practically unexploited field, with so far the only detection taking place in the 1970s\cite{Novick1972,Weisskopf1976,Weisskopf1978a}. In 2001, high-sensitivity soft x-ray polarimetry became possible with the invention of photoelectric polarimeters\cite{Costa2001,Bellazzini2003b}. 

A polarimeter onboard a CubeSat named PolarLight\cite{Feng2019} was launched into a Sun-synchronous orbit on October 29, 2018, as the first flight test of the new technique.  PolarLight is a gas pixel detector capable of measuring x-ray polarisation via electron tracking.  The instrument has a collecting area of only 1.6~cm$^2$, with a collimator to constrain the field of view to 2.3$^\circ$ (full width at half maximum). Therefore, PolarLight can be treated as a miniature x-ray polarimeter built on the basis of a high-sensitivity technique\cite{Bellazzini2013}. 

Since launch, the Crab (here referring to both the nebula and pulsar if not specified) has been the primary target of PolarLight and observed except from early May to early July when the source was too close to the Sun on the sky plane. Here we report the results from observations as of early December, 2019, with a total exposure of about 660~ks for the Crab and 165~ks for the background. Given the Crab spectrum, the polarisation measurement is most sensitive in the energy range of 3.0--4.5 keV.  The background is mainly resulted from charged particles, and some of them can be discriminated by charge morphology.  In 3.0--4.5 keV, valid events that can be used for polarimetry have a mean count rate of about 0.03~\cts\ when observing the Crab, with an average background contamination of about 10\%.  The polarisation fraction (PF) and polarisation angle (PA) are calculated based on the Stokes parameters\cite{Kislat2015}. We adopt the Bayesian approach to perform the point and interval estimates of the intrinsic polarisation. Details about the data reduction and analysis can be found in Methods.

In the 1970s, the Bragg polarimeter onboard OSO-8 measured x-ray polarisation of the Crab at two narrow bands around 2.6 and 5.2 keV, respectively\cite{Weisskopf1976,Weisskopf1978a}. At 2.6 keV, the total Crab emission has ${\rm PF} = 0.157 \pm 0.015$ and ${\rm PA} = 161.1^\circ \pm 2.8^\circ$, and the pulsar-free nebular emission has ${\rm PF} = 0.192 \pm 0.010$ and ${\rm PA} = 156.4^\circ \pm 1.4^\circ$. The results at 5.2 keV are consistent with those at 2.6 keV within errors.  

\begin{figure*}
\centering
\includegraphics[width=0.4\columnwidth]{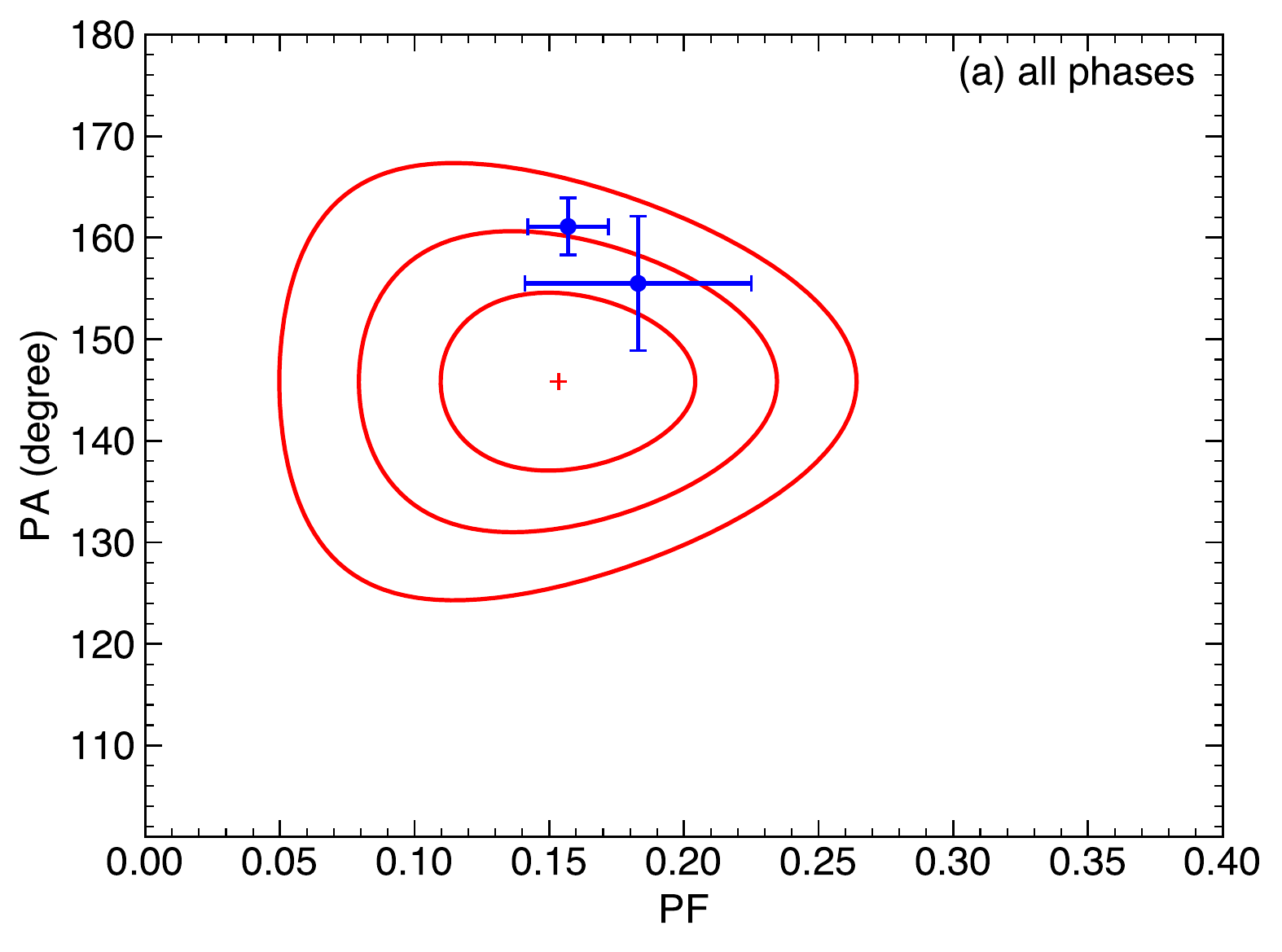}
\includegraphics[width=0.4\columnwidth]{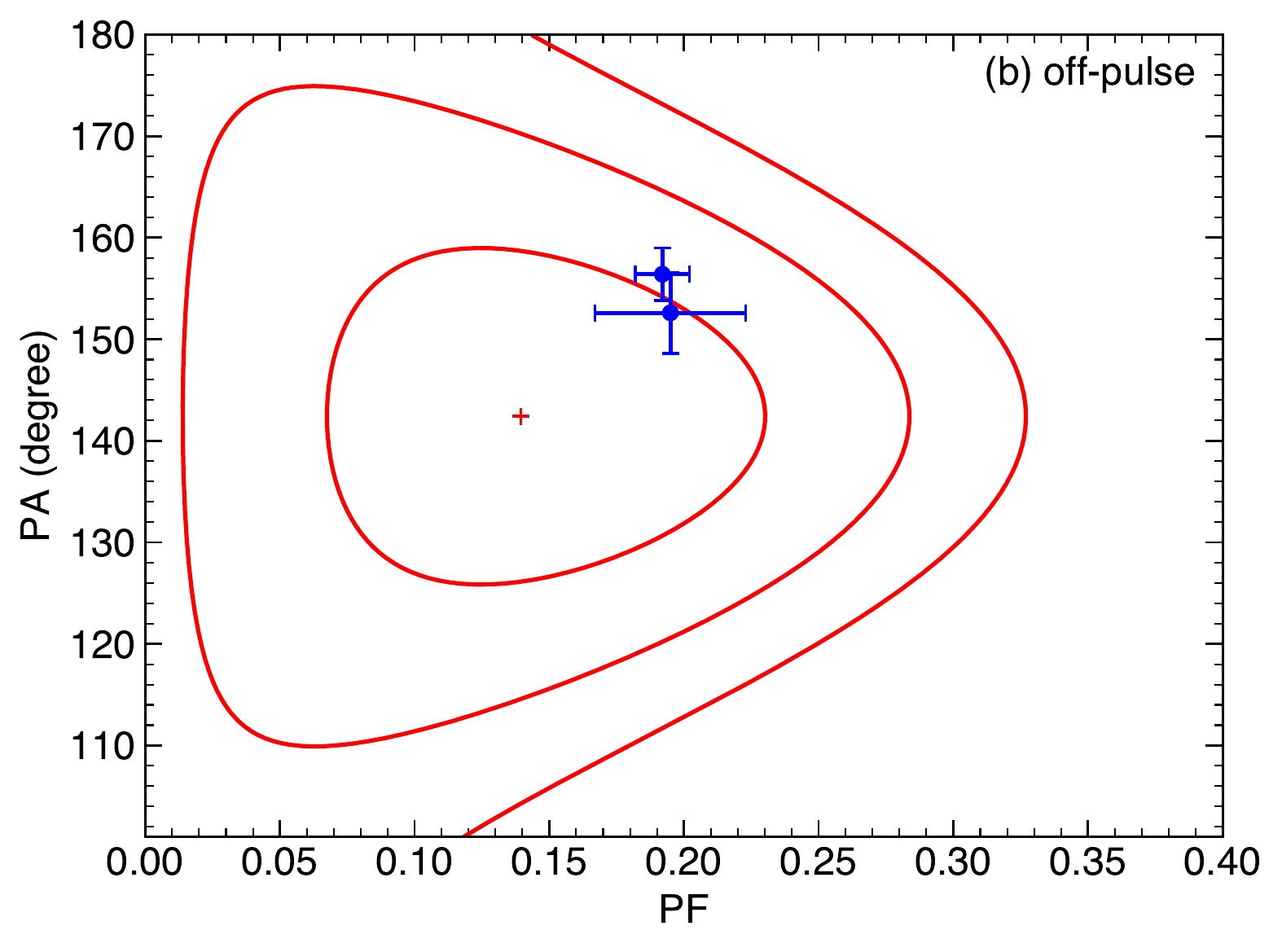}
\includegraphics[width=0.6\columnwidth]{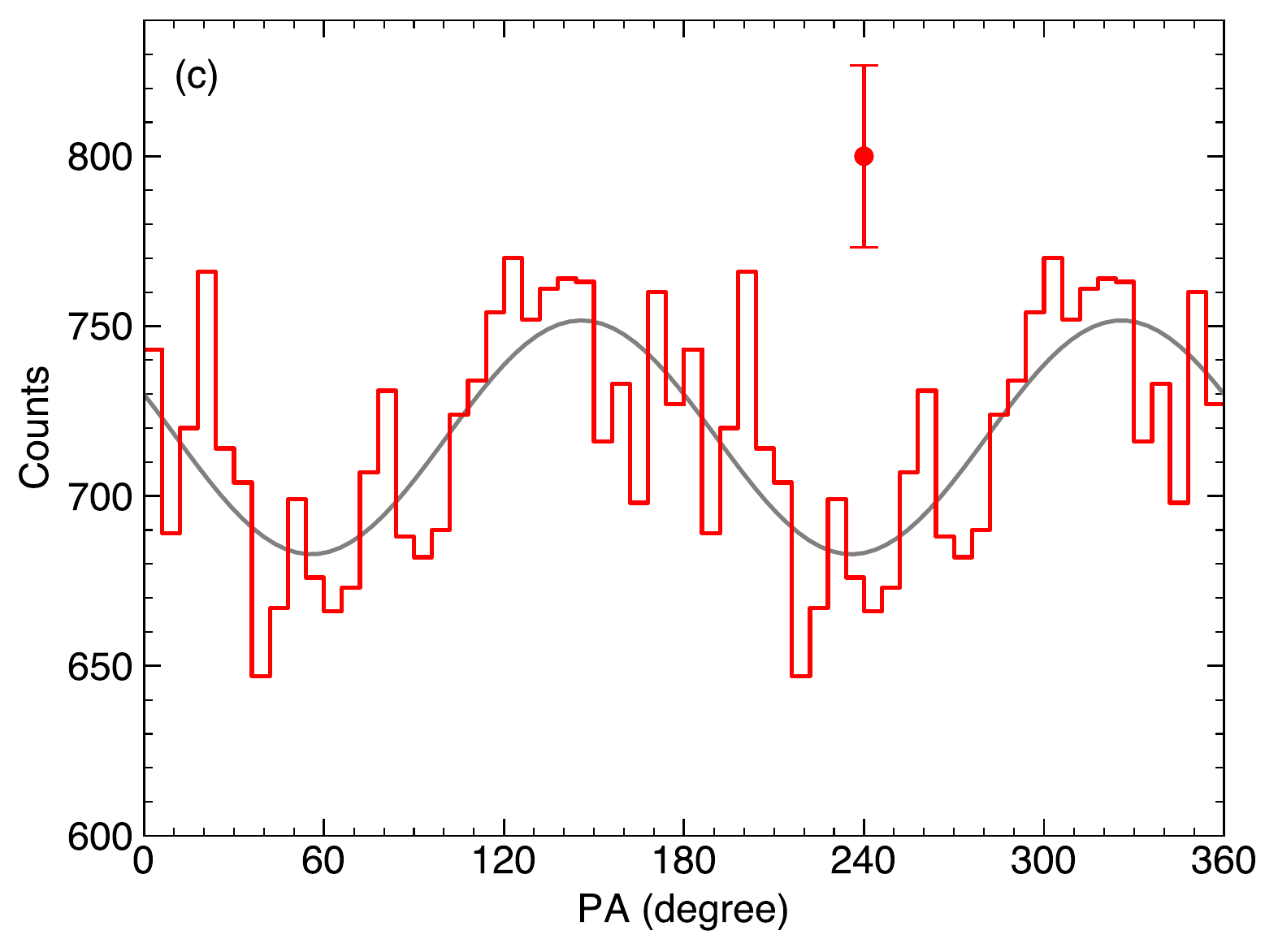} 
\caption{Bayesian posterior distribution of PF and PA derived from PolarLight observations of the Crab in the energy range of  3.0--4.5 keV, for (a) data in all phases and (b) data in the off-pulse phases that include pulsar-free nebular emission. The red plus indicates the point estimate and the contours encircle the 1, 2, and 3$\sigma$ credible intervals.  The blue points and error bars mark the Crab polarisation measured with the polarimeter on OSO-8 at 2.6 keV (one with smaller errors) and 5.2 keV (one with larger errors)\cite{Weisskopf1976,Weisskopf1978a}.  A modulation curve (red) using data in all phases against the model (grey) inferred in this Letter (see parameters in Table~\ref{tab:pol}) is displayed in panel (c) for visual inspection.  The error bar indicates the typical 1$\sigma$ error of the counts in each phase bin.
\label{fig:single}}
\end{figure*}

With PolarLight, time-averaged polarisation measurements of the Crab using all of the data in the energy range of 3.0--4.5 keV is displayed in Fig.~\ref{fig:single}a. The result suggests that the Crab emission in this energy band has an average PF of $0.153_{-0.030}^{+0.031}$ and PA of $145.8^\circ \pm 5.7^\circ$, detected at a significance of 4.7$\sigma$. Our PF is consistent with that obtained with OSO-8, while the PA differs by approximately 2$\sigma$. To compare with the OSO-8 results, polarisation of the pulsar-free nebular emission is shown in Fig.~\ref{fig:single}b, calculated from the off-pulse phase interval (see Extended Data Fig.~\figpsr\ in Methods), which is in line with the definition used in the OSO-8 analysis. Our off-pulse result has a relatively large uncertainty and is consistent with that from OSO-8.  On July 23, 2019, a glitch was detected in the Crab pulsar\cite{Shaw2019}.  To investigate possible phase and time dependence of polarisation, we analyse the data in two phase ranges (on-pulse and off-pulse) and two epochs (before and after the glitch), respectively.  The results are listed in Table~\ref{tab:pol}.  

\begin{figure*}
\centering
\includegraphics[width=0.49\columnwidth]{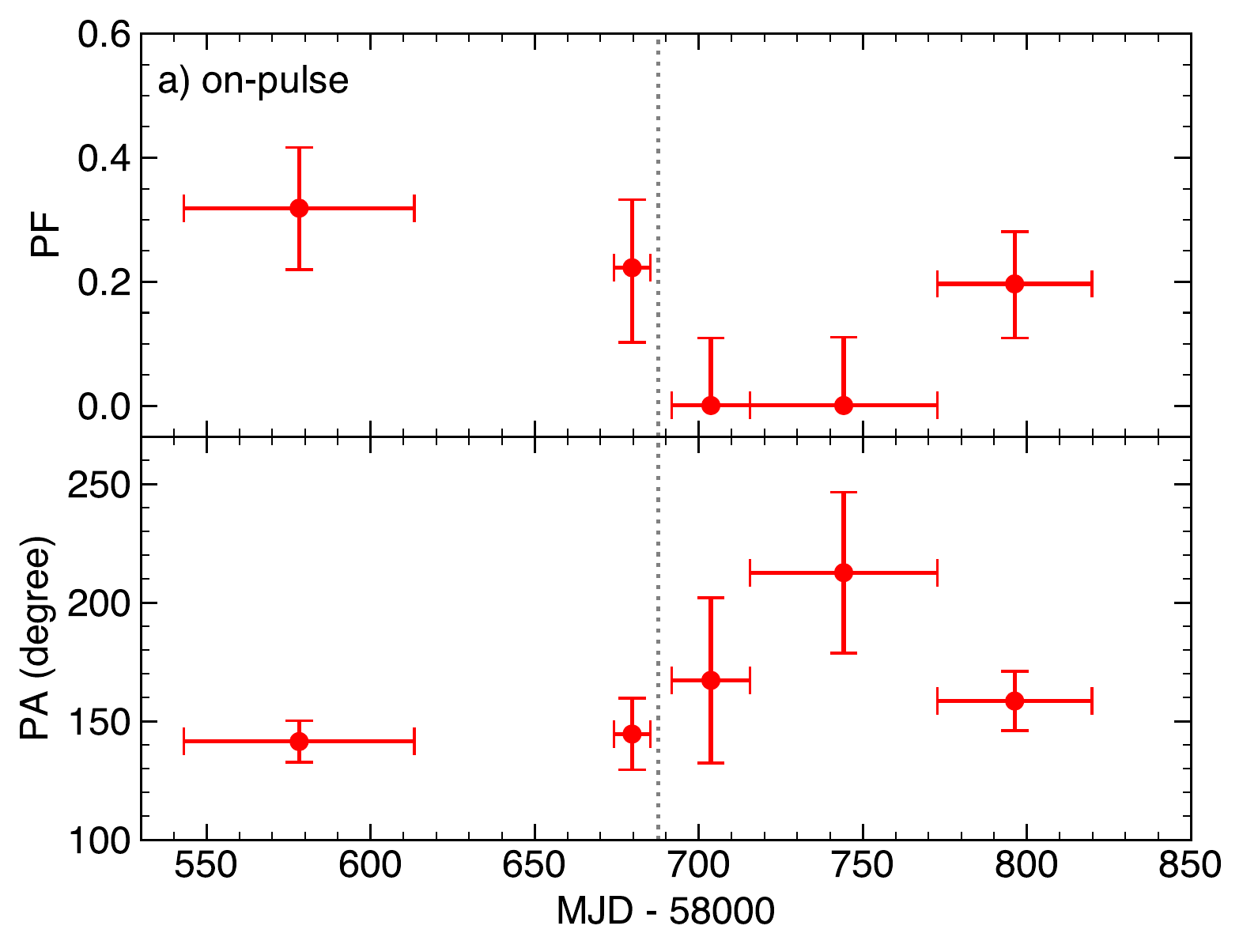}
\includegraphics[width=0.49\columnwidth]{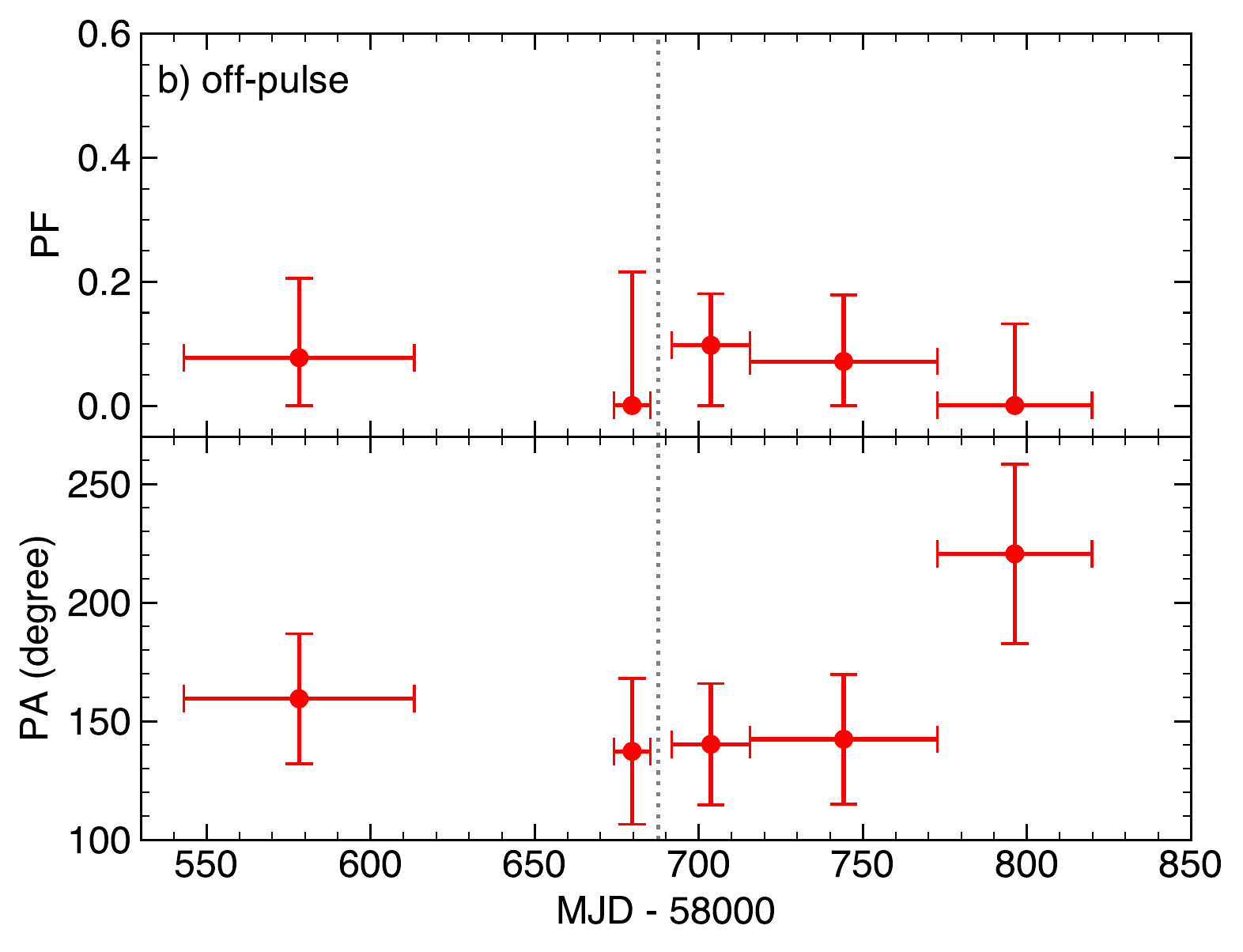}
\caption{The Crab x-ray polarisation in the 3.0--4.5 keV band as a function of time, respectively, for events in (a) on-pulse phases and (b) off-pulse phases.  The first bin includes all observations in the first observing window (2019 May and before). As observations could not be scheduled efficiently in that epoch, it spans over a large time interval. The second bin includes observations in the second window (2019 July and after) but before the Crab glitch on July 23, 2019, marked by a vertical dotted line.  The following bins are from the rest observations in the second window equally divided by number of photons. The net effective exposure in each bin is 90,  98,  168,  158,  and 145~ks, respectively. The horizontal bar indicates the time span of the measurements. The vertical error bar indicates the 68\% credible interval of PF or PA.  We note that PA is wrapped between 0 and 180 degrees by definition, but to avoid a visual impression for a large PA variation, here we have mapped some PAs around 30--40 degrees to 210--220 degrees. The on-pulse polarisation is observed to vary after the pulsar glitch, while the nebular polarisation remains constant. 
\label{fig:time}}
\end{figure*}

The most interesting finding is that a time variation of polarisation is detected for emission during the on-pulse phases. The PF is found to decrease from $0.288_{-0.073}^{+0.071}$ before the glitch to $0.101_{-0.051}^{+0.047}$ after the glitch.  We divide the data into smaller time bins and calculate polarisations in the on-pulse and off-pulse phases, respectively (Fig.~\ref{fig:time}). The on-pulse polarisation exhibited an abrupt change after the glitch, while the off-pulse polarisation remained constant.  The Bayes factor is adopted to test the significance of a constant PF against a PF decrease over different time intervals after the glitch (see Methods).  A PF decrease after the glitch is favoured with a strong evidence using data after 30--100 days of the glitch (Fig.~\ref{fig:bftime}).  Within $\sim$30~days after the glitch, the statistic is not high enough to claim any statement. After $\sim$100 days of the glitch, the PF started to recover leading to a weaker evidence, which is consistent with the polarisation lightcurve shown in Fig.~\ref{fig:time}.  A change (either decrease or increase) in PF after the glitch is found to have a substantial evidence, also shown in Fig.~\ref{fig:bftime}.  As an independent test, the posterior distribution of polarisation derived from data before the glitch is compared with that within 100 days after the glitch (see Methods). The two measurements are not consistent with each other at a 3$\sigma$ level, suggesting that the above conclusion is valid.  In addition, a bootstrap test produces a consistent result that the decrease is evident at a significance of 3$\sigma$ (see Methods).  We emphasise that the choice of 100 days after the glitch is not subjective; choosing any time from 30--100 days after the glitch results in the same conclusion, as shown in Fig.~\ref{fig:bftime}.

The rate of directly measured background, as well as the rate of events rejected by particle discrimination, is found to be constant on timescales longer than the orbital period (see Methods). No polarisation modulation in background is detected. Taking into account the upper limit of polarisation and flux contribution, the background is unable to account for the observed change in PF (see Methods). The contribution of background is the same to both on-pulse and off-pulse results. Thus, the variation can not be a result of change in background.  The instrumental systematic error on the PF is below 1\% (see Methods). Laboratory tests indicate that the modulation factor varies no more than 4\% in response to the change of gain (see Methods). The uncertainty of modulation factor due to inaccurate gain calibration is less than 5\% (see Methods). Similarly, a change in the modulation factor will affect results in all phases and can be ruled out.  To conclude, we claim strong evidence for a decrease in PF after the glitch of the Crab pulsar, and rule out the possibility that this is due to background or calibration systematics.

The results suggest that the variation is more likely associated with the pulsar, because it is more significant in the on-pulse phases. The pulsar occupies roughly 7--8\% of the total Crab emission in the energy band of 3.0--4.5 keV\cite{Thomas1975}, or $\sim$12\% in our on-pulse phase interval. Such a flux fraction is possible to produce the observed variation only if the pulsar PF is high. In that case, the observed PF decrease could be due to a large PA variation from the pulsar emission after the glitch. We try to estimate the pure pulsar polarisation by subtracting the off-pulse polarisation from on-pulse signals. However, no useful constraint can be made with the current data.  A high degree of polarisation during the on-pulse phases challenges nearly all of the pulsar emission models, which predict either a swing of PA with phases that leads to a low PF on average or a considerably low PF during the on-pulse phases\cite{Dyks2004a,Takata2007,Harding2017}. A change of the nebular polarisation is another possibility, which cannot be ruled out by the currently available data.

Glitches are thought to be a result of catastrophic superfluid vortex unpinning\cite{Baym1969,Anderson1975,Alpar1993}. In the vortex model, the sudden angular momentum transfer between the faster rotating superfluid interior and the crust results in an abrupt spin-up. The motion of the core superfluid vortices may alter the core magnetic flux tubes\cite{Ruderman1998}. The sudden spin-up of the crust may cause a change in the configuration of magnetic fields threaded in it, and consequently in the co-rotating magnetosphere. For Crab-like young pulsars, starquakes may trigger the glitch\cite{Alpar1993} and lead to a natural change in the magnetosphere, especially if the crust cracks in the polar region. The timescale of the polarisation variation that we observed here (tens of days) for the Crab is much longer than that in the radio band observed in the Vela pulsar (a few seconds)\cite{Palfreyman2018}, but is consistent with the post-glitch recovery timescale of the Crab pulsar\cite{Wong2001}.  For rotation-powered pulsars, the emission regions of X-rays and radio may be different\cite{Cheng1986a,Muslimov2004,Kalapotharakos2012}. The mechanism that drives the glitches could also be different for Vela and Crab\cite{Alpar1996}.  As x-rays can be emitted in a large range of altitudes all the way from regions close to the surface to the light cylinder or even beyond, it is very likely that a small change in the magnetosphere could result in a large change in the observed PA via the leverage effect.  

This experiment demonstrates that a CubeSat with a tiny detector for a dedicated science goal can yield valuable data. Such space projects with a relatively low cost and short duration are also ideal platforms for student training.  The successful operation of PolarLight indicates that the window of x-ray polarimetry in the few keV energy range has been re-opened after 40 years. Future missions like the Imaging X-ray Polarimetry Explorer (IXPE)\cite{Weisskopf2016} and enhanced X-ray Timing and Polarimetry (eXTP)\cite{Zhang2019} will have a much higher sensitivity and can perform in-depth studies of this topic. 

\section*{References}

%%%%%%%%%%%%%%%%%%%%%%%%%%%%%%%%%%%%
% \bibliography{xpol,book}

%%%%%%%%%%%%%%%%%%%%%%%%%%%%%%%%%%%%

\begin{addendum}
 \item We thank Hsiang-Kuang Chang, Jumpei Takata, Mingyu Ge, and Jeremy Heyl for helpful discussions. HF acknowledges funding support from the National Natural Science Foundation of China under the grant Nos.\ 11633003 \& 11821303, and the National Key R\&D Project (grants Nos.\ 2018YFA0404502 \& 2016YFA040080X). 
 \item[Author contributions]  H.F.\ is the PI of PolarLight and led the project. H.L.\ and X.L.\ conducted the daily operation of the CubeSat and had a major contribution to the data analysis. R.B.\ led the development of the GPD. E.C., P.S., and F.M.\ participated in the discussion, and E.C.\ had a special contribution to the initiate of the project.  J.H.\ performed the simulation and modelling of the in-orbit background. Q.W., W.J., M.M., D.Y., L.B., S.C., H.N., A.J., J.Y., G.J., M.Z., P.A., A.B., L.L., C.S., G.S., and M.P.\ contributed to the development of the payload instrument. W.W.\ and R.X.\ participated in the interpretation of the results.
 \item[Competing Interests] The authors declare that they have no competing financial interests.
 \item[Correspondence] Correspondence and requests for materials should be addressed to H.F.\ (email: hfeng@tsinghua.edu.cn).
\end{addendum}

\clearpage 

%%%%%%%%%%%%%%%%%%%%%%%%%%%%%%%%%%%%
\renewcommand{\figurename}{Fig.}

\begin{figure*}
\centering
\includegraphics[width=0.49\columnwidth]{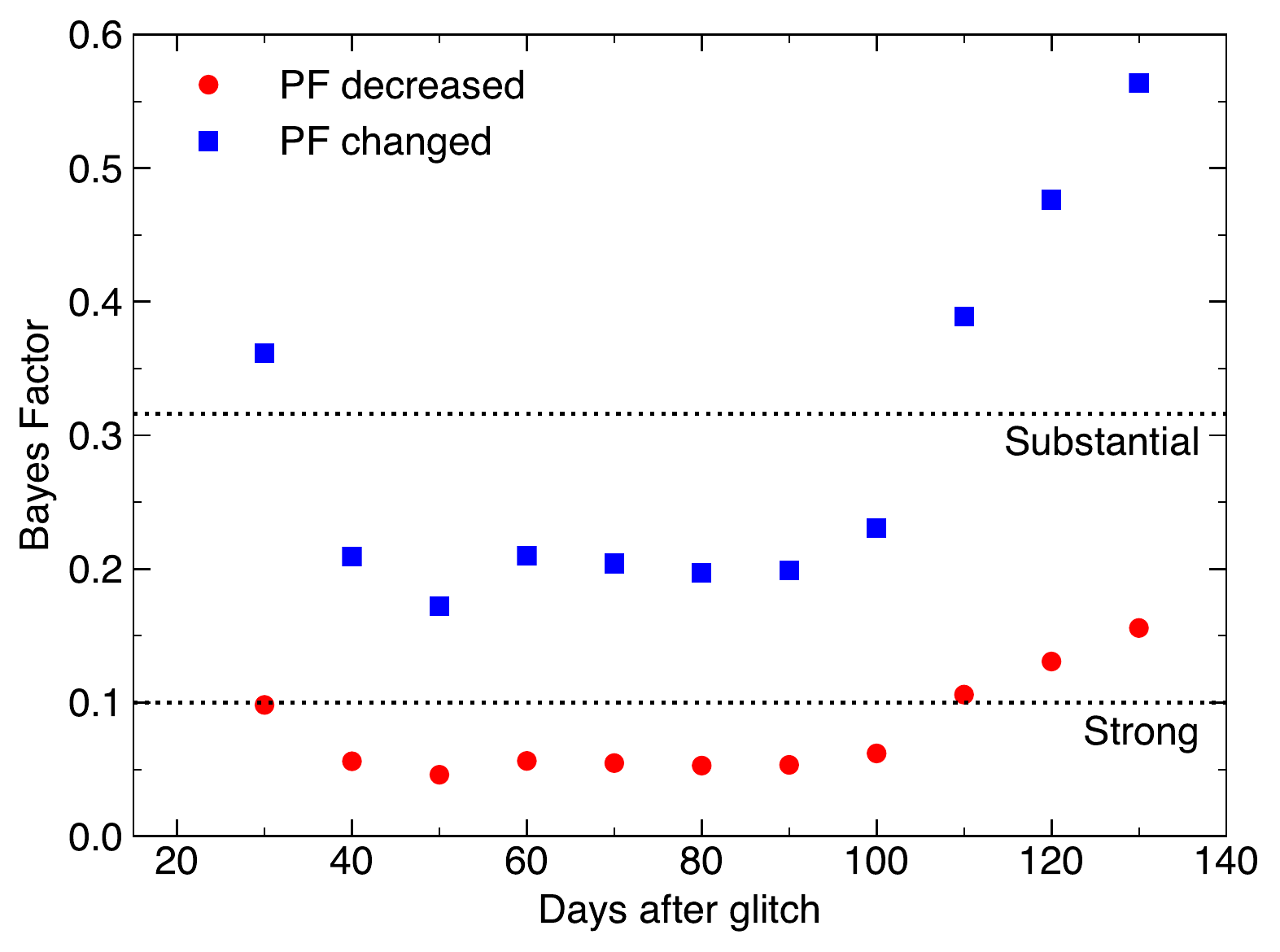}
\caption{Bayes factor to test the model ($\mathcal{M}_0$) that the PF remained constant against the model ($\mathcal{M}_1$) that there was a decrease (red) or change (blue; either decrease or increase) in PF after the glitch. As the variation may be a one-time transient behaviour, the test is done with data until some days after the glitch, which is indicated as the x-axis.  If the Bayes factor is lower than $10^{-1/2}$, the evidence against $\mathcal{M}_0$ is regarded as substantial; if it is lower than $10^{-1}$, the evidence against $M_0$ is strong\cite{Jeffreys1961}. The results suggest that there is a strong evidence that the PF decreased after the glitch and lasted about 100 days, or a substantial evidence that there was a change (either decrease or increase) after the glitch. 
\label{fig:bftime}}
\end{figure*}

\clearpage

%%%%%%%%%%%%%%%%%%%%%%%%%%%%%%%%%%%%
\begin{table*}
\centering
\footnotesize
\begin{threeparttable}
\caption{Phase and time dependent polarisation measurements of the Crab in the energy range of 3.0--4.5 keV with PolarLight.}
\label{tab:pol}
\begin{tabular}{cccc}
\noalign{\bigskip}\hline\hline\noalign{\smallskip}
Phase$^\dagger$ & Epoch$^\ddagger$ & PF & PA \\
& & & (degree) \\
\noalign{\smallskip}\hline\noalign{\smallskip}
all & all & $0.153_{-0.030}^{+0.031}$ & $145.8 \pm 5.7$ \\
\noalign{\smallskip}\hline\noalign{\smallskip}
off-pulse & all & $0.140_{-0.054}^{+0.052}$ & $142.4 \pm 11.0$ \\
on-pulse & all & $0.158_{-0.039}^{+0.039}$ & $147.6 \pm 7.0$ \\
\noalign{\smallskip}\hline\noalign{\smallskip}
all & before glitch & $0.243_{-0.057}^{+0.057}$ & $144.5 \pm 6.7$ \\
all & after glitch & $0.113_{-0.038}^{+0.037}$ & $146.9 \pm 9.6$ \\
\noalign{\smallskip}\hline\noalign{\smallskip}
off-pulse & before glitch & $0.137_{-0.110}^{+0.076}$ & $149.9 \pm 21.0$ \\
off-pulse & after glitch & $0.127_{-0.067}^{+0.061}$ & $138.7 \pm 15.1$ \\
on-pulse & before glitch & $0.288_{-0.073}^{+0.071}$ & $142.7 \pm 7.2$ \\
on-pulse & after glitch & $0.101_{-0.051}^{+0.047}$ & $153.0 \pm 14.4$ \\
\noalign{\smallskip}\hline\noalign{\smallskip}
\end{tabular}
\begin{tablenotes}
 \item[$^\dagger$] For emission in all phases, the on-pulse phase interval that contains both pulsar and nebula emission, or the off-pulse interval that is free of pulsar emission. See Methods for definition. 
 \item[$^\ddagger$] Data in all times, before or after the Crab glitch on July 23, 2019.
 \item[] Errors are quoted as 68\% credible intervals.
\end{tablenotes}
\end{threeparttable}
\end{table*} 
%%%%%%%%%%%%%%%%%%%%%%%%%%%%%%%%%%%%

\clearpage

\begin{methods}

\subsection{Observations.}
PolarLight flies in a high-inclination low-Earth orbit, where the charged particle flux is high in the two polar regions and the South Atlantic Anomaly region. The high voltage power supply of the detector needs be powered off  in these regions.  Taking into account the Earth occultation and other constraints (e.g.,  by the star tracker), the total duration in a day that the target is observable is roughly 200 minutes.  We note that some orbits are not used if the continuous observable time is not long enough to power on and off the high voltage power supply. The first observation of the Crab took place on March 1, 2019. About one month later, regular observations could be scheduled in an efficient way. The Crab has been observed routinely since then. From May 11 to July 10 in 2019, no observation was scheduled for the Crab because the source was close to the Sun on the sky plane.  In this letter, observations as of December 3, 2019 are included. After that date, there were several major upgrades of the onboard computer and science observations were paused for a while. 

\subsection{Gain calibration.}
The detector gain is calibrated by comparing a simulated Crab spectrum\cite{Kirsch2005}, an absorbed power-law spectrum with an absorption column density of $4.5 \times 10^{21}$~cm$^{-2}$ and a photon index of 2.07, with the observed pulse height spectrum. The background spectra are obtained via observations of the blank sky and Earth atmosphere. The two background spectra show no difference as charged particles are the main cause of the background, which is confirmed by simulations.  The pulse height spectrum of the background can be described by an exponentially cutoff power-law function, with parameters found from pure background observations. In the energy calibration, we fit the observed spectrum during Crab observations with the sum of the two components, with the spectral shapes fixed but the normalisations and pulse height to energy relation as free parameters. Because the two components dominate at different energy ranges (see Extended Data Fig.~\ref{fig:spec}), they can be readily decoupled. With the gain calibrated in such a way, the peak position of the energy spectrum at different epochs has a mean of 2.16~keV with a standard deviation of 0.012~keV, suggesting that the uncertainty for the gain calibration is on the level of 1\%. The track eccentricity is a function of energy, and its distribution in 3.0--4.5 keV shows a consistent shape along time, allowing us to constrain the gain uncertainty to be better than 5\%. Even taking the more conservative estimate, the gain uncertainty will have no effect on the accuracy of the results. 

\subsection{Data reduction.}
The science data for each event include the arrival time and track image.  With the track image, the photon energy and emission angle of the photoelectron on the detector plane can be inferred.  Charged particles will also deposit energies in the detector, but usually leave a long, straight track behind, with multiple charge islands. To distinguish them, we select x-ray events if the image has a diagonal no more than 70 pixels, an eccentricity not higher than 50, and only one isolated charge island. This step is called particle discrimination.  Events near the edge of the detector (outside the central 14~mm $\times$ 14~mm region) are not used as they may have a partial deposit of charges. Then, to remove spurious modulation at a period of 60$^\circ$ due to the hexagonal pixels\cite{Muleri2010}, we discard events with charges spreading on less than 58 pixels in polarisation analysis.  The choices of these criteria and parameters are justified by simulations and laboratory tests at 3.74~keV using 45$^\circ$ diffraction with the Al crystal.  Given the modulation factor $\mu$, measured counts from the Crab $N$, and background fraction $f_{\rm b}$, we calculate the quality factor $\left[\equiv \mu \sqrt{N} \left(1 - f_{\rm b}\right)\right]$ of the detector as a function of energy in Extended Data Fig.~\ref{fig:quality}.  The quality factor is an indicator of the sensitivity of polarimetry. As one can see, 3.0--4.5 keV is the most sensitive energy range for polarimetry of the Crab. Inclusion of data down to 2 keV or up to 5 keV does not improve the polarimetric signal to noise ratio. In the energy range of 3.0--4.5 keV, the average modulation factor\cite{Feng2019} is 0.35, and the background flux fraction is 10\%.

\subsection{Phase definition.}
The photon arrival time is converted to the solar system barycentre using the JPL DE430 ephemeris\cite{Folkner2014}.  The Jodrell Bank ephemeris for the Crab pulsar\cite{Lyne1993} is used to fold the pulse profile, displayed in Extended Data Fig.~\ref{fig:psr}. To be in line with the definition used for the OSO-8 analysis\cite{Silver1978}, the phase interval from 17~ms to 30~ms after the primary pulse peak is defined as the off-pulse interval, while the rest is defined as the on-pulse interval. In our case, the on-pulse interval corresponds to a phase range from 0 to 0.62.

\subsection{Polarisation measurement.}
The polarisation is calculated based on the Stokes parameters\cite{Kislat2015,Mikhalev2018}. Here we use the subscript ``r'' to denote reconstructed values and subscript ``0'' to denote intrinsic values. Given the position angle ($\phi$) of each photoelectron, the source and background counts ($S$ and $B$), and the average modulation factor ($\mu$) in the energy band, the normalised Stokes parameters are
\begin{eqnarray}
Q_{\rm r} &=& \frac{1}{S}\sum_{i = 1}^{S+B} \cos\left(2\phi_i\right) \; , \\
U_{\rm r} &=& \frac{1}{S}\sum_{i = 1}^{S+B} \sin\left(2\phi_i\right) \; .
\end{eqnarray}
 The reconstructed PF ($p_{\rm r}$) and PA ($\psi_{\rm r}$) are
 \begin{eqnarray}
p_{\rm r} &=& \frac{2}{\mu}\sqrt{Q_{\rm r}^2 + U_{\rm r}^2} \; , \\
\psi_{\rm r} &=& \frac{1}{2} \arctan\frac{U_{\rm r}}{Q_{\rm r}} \; .
\end{eqnarray}
Due to the positive-definite nature of polarisation measurement, the estimate of PF will introduce a bias. The situation is getting worse towards low statistics, e.g., when one tries to divide the data into smaller time or phase bins. The bias could be corrected by using the Bayesian approach\cite{Maier2014,Mikhalev2018}, which is adopted to estimate the intrinsic parameters in this work.  The prior distribution of $p_0$ is assumed to be uniform between 0 and 1, and that of $\psi_0$ is assumed to be uniform between 0 and $\pi$. The probability of measuring $p_{\rm r}$ and $\psi_{\rm r}$ given the intrinsic $p_0$ and $\psi_0$ is
\begin{equation}
\rho\left(p_{\rm r}, \psi_{\rm r} \mid p_0, \psi_0\right) = \frac{p_r}{\pi \sigma^2} \exp\left(-\frac{p_{\rm r}^2 + p_0^2 - 2p_{\rm r}p_0 \cos\left(2\psi_{\rm r} - 2\psi_0\right)}{2\sigma^2}\right) \; ,
\label{eq:like}
\end{equation}
where $\sigma = \sqrt{2(S+B)} / (\mu S)$.  The posterior distribution is written as
\begin{equation}
\rho\left(p_0, \psi_0 \mid p_{\rm r}, \psi_{\rm r}\right) =  \frac{\rho\left(p_0, \psi_0\right) \rho\left(p_{\rm r}, \psi_{\rm r} \mid p_0, \psi_0\right)}{\iint \rho\left(p_0, \psi_0\right) \rho\left(p_{\rm r}, \psi_{\rm r} \mid p_0, \psi_0\right) \mathrm{d}p_0 \mathrm{d}\psi_0} \; .
\label{eq:post}
\end{equation}
To estimate $p_0$ and $\psi_0$, we use the marginalised posterior distributions. For $p_0$, the maximum a posteriori (MAP) of the marginalised posterior distribution is adopted for point estimate. $\psi_{\rm r}$ is an unbiased estimate of $\psi_0$ and is adopted for point estimate directly. The credible interval (the region of the highest posterior density) from the marginalised posterior distribution given a probability (68\% or 90\%; always specified in the text) is quoted as the error range. We note that the credible interval is non-symmetric for $p_0$ due to the nature of Rice distribution, but symmetric for $\psi_0$\cite{Maier2014}.  

\subsection{Test of a decrease/change in PF with the Bayes factor.} 
Two models are defined, with $\mathcal{M}_0$ referring to a constant PF and $\mathcal{M}_1$ referring to a PF decrease after the glitch. The Bayes factor BF$_{01}$, $= P\left(\mathcal{D} \mid \mathcal{M}_0\right) / P\left(\mathcal{D} \mid \mathcal{M}_1\right)$, is calculated to quantify how much $\mathcal{M}_0$ is favoured over $\mathcal{M}_1$ from the data $\mathcal{D}$.  The likelihood is calculated using Monte-Carlo integration\cite{Chauvin2018a} as
\begin{equation}
P\left(\mathcal{D} \mid \mathcal{M}_i\right) \approx \frac{1}{N} \sum_{j = 1}^{N} P\left(\mathcal{D} \mid \theta_j, \mathcal{M}_i\right) \quad (i = 0,1),
\end{equation}
where $\theta$ is the assembly of model parameters that follow $P\left(\theta \mid M_i\right)$, and $N$ is the Monte-Carlo sample size and should be sufficiently large.  The data include all the observations before the glitch and those within a certain range of days after the glitch.  $\theta$ includes two intrinsic PFs, $p_{0, {\rm a}}$ before the glitch and $p_{0, {\rm b}}$ after the glitch, with $p_{0, {\rm a}} = p_{0, {\rm b}}$ for $\mathcal{M}_0$ and $p_{0, {\rm a}} > p_{0, {\rm b}}$ for $\mathcal{M}_1$.  They are randomly sampled from the allowed parameter space. Then, with $S$ and $B$ in the two epochs and $\mu$, the probability of data given the parameters can be computed as
\begin{equation}
P\left(\mathcal{D} \mid \theta, \mathcal{M}\right) = P\left(p_{\rm r, a} \mid p_{0, {\rm a}}\right)  P\left(p_{\rm r, b} \mid p_{0, {\rm b}}\right)  \; ,
\end{equation}
where $P\left(p_{\rm r} \mid p_0\right)$ is the likelihood of $p_{\rm r}$ given $p_0$, a Rice distribution that can be obtained by integrating Equation (\ref{eq:like}) over $\psi$. In the case to test against a change (either increase or decrease) in PF after the glitch, one simply needs to modify $\mathcal{M}_1$ to have two independent intrinsic PFs, $p_{0, {\rm a}}$ and $p_{0, {\rm b}}$, respectively before and after the glitch.

\subsection{Other evidence for a decrease in polarisation after the glitch.}
Here we compare two samples. Sample A contains data before the glitch and Sample B contains data within 100 days after the glitch. The end date of 100 days after the glitch is not a subjective choice; the same conclusion remains if one chooses any date from $\sim$30 days to $\sim$100 days after the glitch (see Fig.~\ref{fig:bftime}). The posterior distributions of polarisation with the two samples are plotted in Extended Data Fig.~\ref{fig:two_cont}. Each measurement is not consistent with the other at a 3$\sigma$ level.  We also do bootstrap to test if one result can be seen in the other sample. Given the number of events in Sample B, we resample with replacement (each event has an equal probability of being selected) the events in sample A and measure the polarisation for 100,000 times, and find that there is 1 time in which the PF is lower than measured from the data in Sample B, suggesting that the time variation is detected at a significance of 4.4$\sigma$. Alternatively, resampling Sample B with the number of events in Sample A, there are 299 times in which the PF is higher than measured in Sample A, corresponding to a significance of 3$\sigma$. Considering that Sample A has 3969 events and Sample B has 7671 events, the second approach is more appropriate and adopted. To conclude, the two independent means both suggest that a decrease is evident at a 3$\sigma$ level.

% 1. background rate; 2. background modulation; 3. residual modulation; 4. energy calibration; 5. mu vs. gain
\subsection{Background and possible systematics.}
The measured x-ray count rate in the 3.0--4.5 keV band when observing the Crab and background regions is displayed in Extended Data Fig.~\ref{fig:lc}. As one can see, there is no obvious change of the background rate after the glitch.  If a PF change with a factor of nearly 3 is caused by a change of background rate, the background fraction is required to vary from 10\% to 70\%, which is certainly not observed. When observing the background regions, the satellite is operated in the magnetic control mode and the star track is usually not valid. Thus, we can only calculate the polarisation of background in the detector plane rather than in the sky plane. In fact, instrument rotation during observations will lower the background modulation, if any, on the sky plane.  The total number of photons in 3.0--4.5 keV collected in the background regions after particle discrimination is only 680.   The PF in the background data can not be detected, with a 90\% upper limit of 0.28, which, along with a flux fraction of 10\%, is insufficient to account for the observed change in PF. The residual modulation of this type of detector is below 1\% averaged over the whole detector plane\cite{Li2015}. For the PolarLight flight model, due to a tight schedule, we did not calibrate its residual modulation because the statistical limit is well above the systematic limit.  The gain uniformity test with a $^{55}$Fe source measured $\rm{PF} = 0.009_{-0.009}^{+0.007}$  (90\%) with about $2.3 \times 10^4$ photons. Thus, the residual modulation of the detector must be low, and it should be a constant effect and can not account for the variation. Another possibility of systematics is that the modulation factor was misestimated after the glitch. This is possible if the gain calibration is not accurate. As mentioned above, the uncertainty of gain calibration is roughly 1\%, or less than 5\% conservatively.  This yields an uncertainty of 5\% in $\mu$, which is not sufficient to account for the observed variation in PF.  Also, the modulation factor is found to vary no more than 4\% at 3.74 keV (measured with the Al crystal) in a gain range that covers the observed range in the orbit.  This, again, is unable to account for the observed change in PF.  We want to emphasise that all these possible systematics or background effects will have the same effect on both on-pulse and off-pulse results, and can thus be ruled out.  

\end{methods}

\begin{addendum}
 \item[Data availability] The datasets generated and analysed in this study are available from the corresponding author on reasonable request.
\end{addendum}

%%%%%%%%%%%%%%%%%%%%%%%%%%%%%%%%%%%%
\section*{Additional References}

%%%%%%%%%%%%%%%%%%%%%%%%%%%%%%%%%%%%%%%%%%%%

\clearpage

\renewcommand{\figurename}{Extended Data Figure}
\setcounter{figure}{0}

\begin{figure*} 
\centering
\includegraphics[width=0.5\columnwidth]{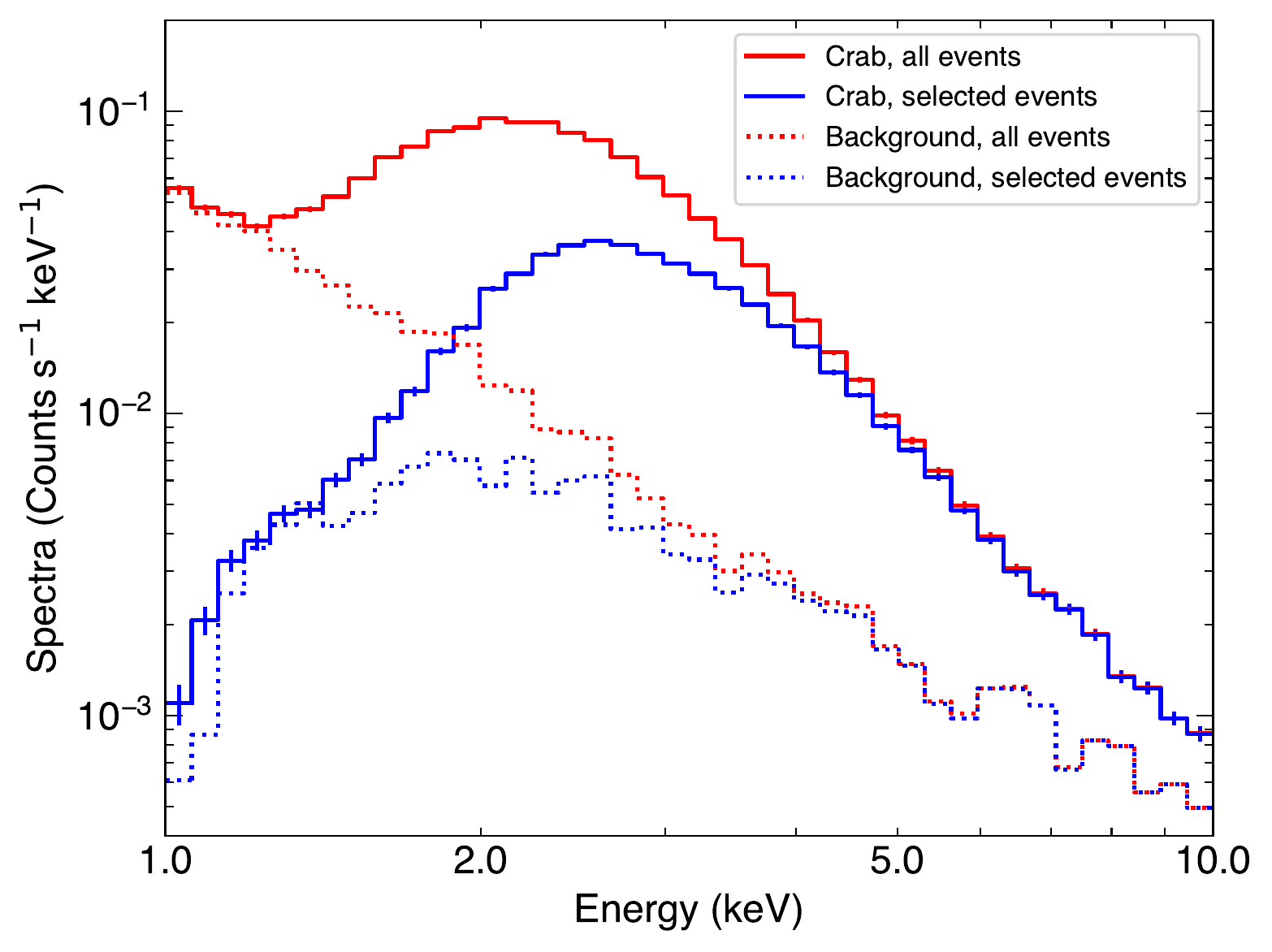}
\caption{Energy spectra of the Crab (solid) and background (dotted). The background spectra are obtained by observing source-free regions. The red spectra are constructed using all x-ray events passing from particle discrimination and the blue ones consist of events used for polarimetry (with one more criterion on the number of fired pixels).  Errors of 1$\sigma$ are shown on the two Crab spectra. We note that the background events shown in the plot are mainly due to charged particles but can not be distinguished by particle discrimination. A discussion on the time variation and modulation of the background can be found in Methods.
\label{fig:spec}}
\end{figure*}

%\clearpage

\begin{figure*}
\centering
\includegraphics[width=0.5\columnwidth]{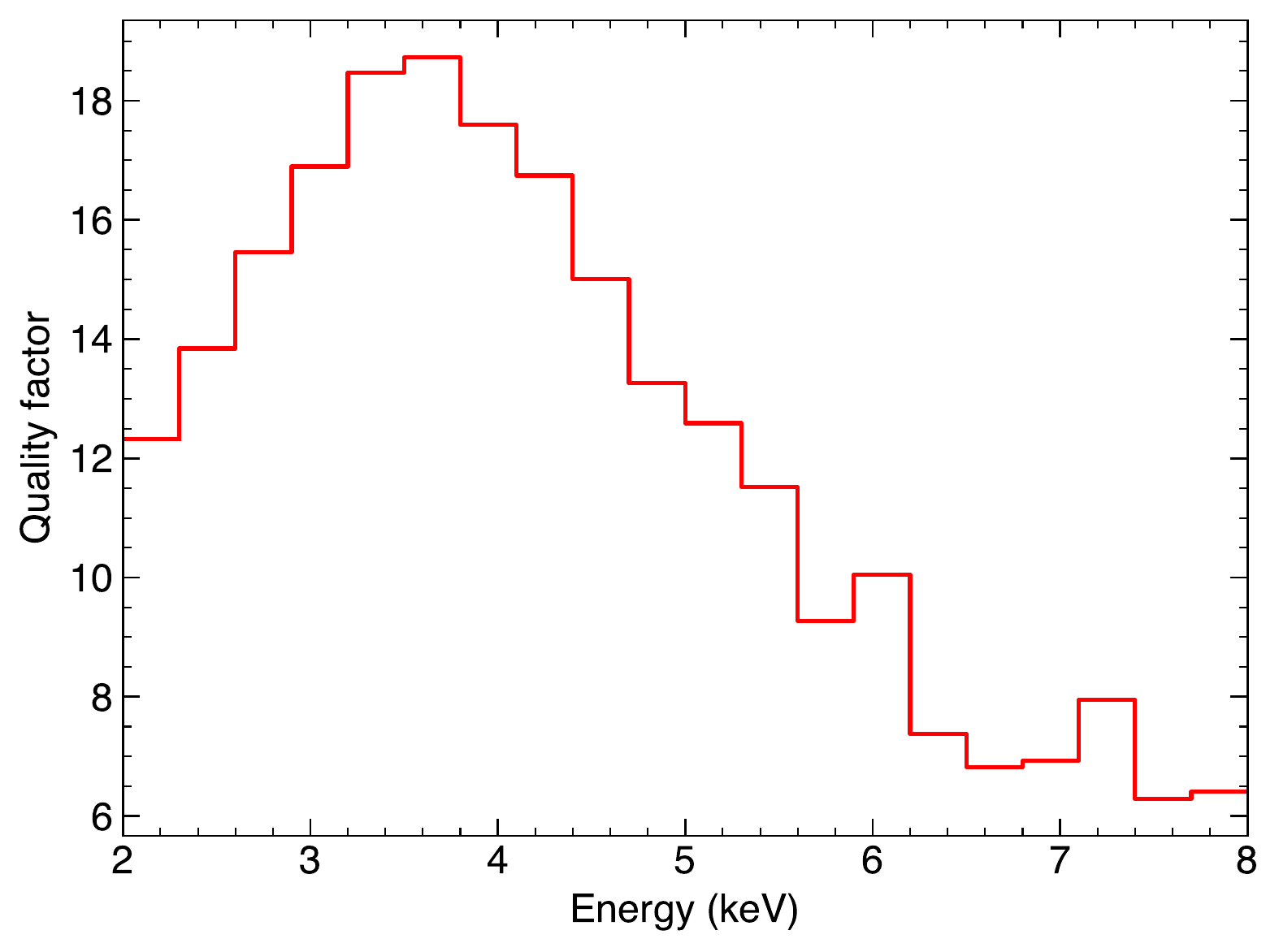}
\caption{Polarisation quality factor of PolarLight when observing the Crab.  
\label{fig:quality}}
\end{figure*}

%\clearpage

\begin{figure*}
\centering
\includegraphics[width=0.5\columnwidth]{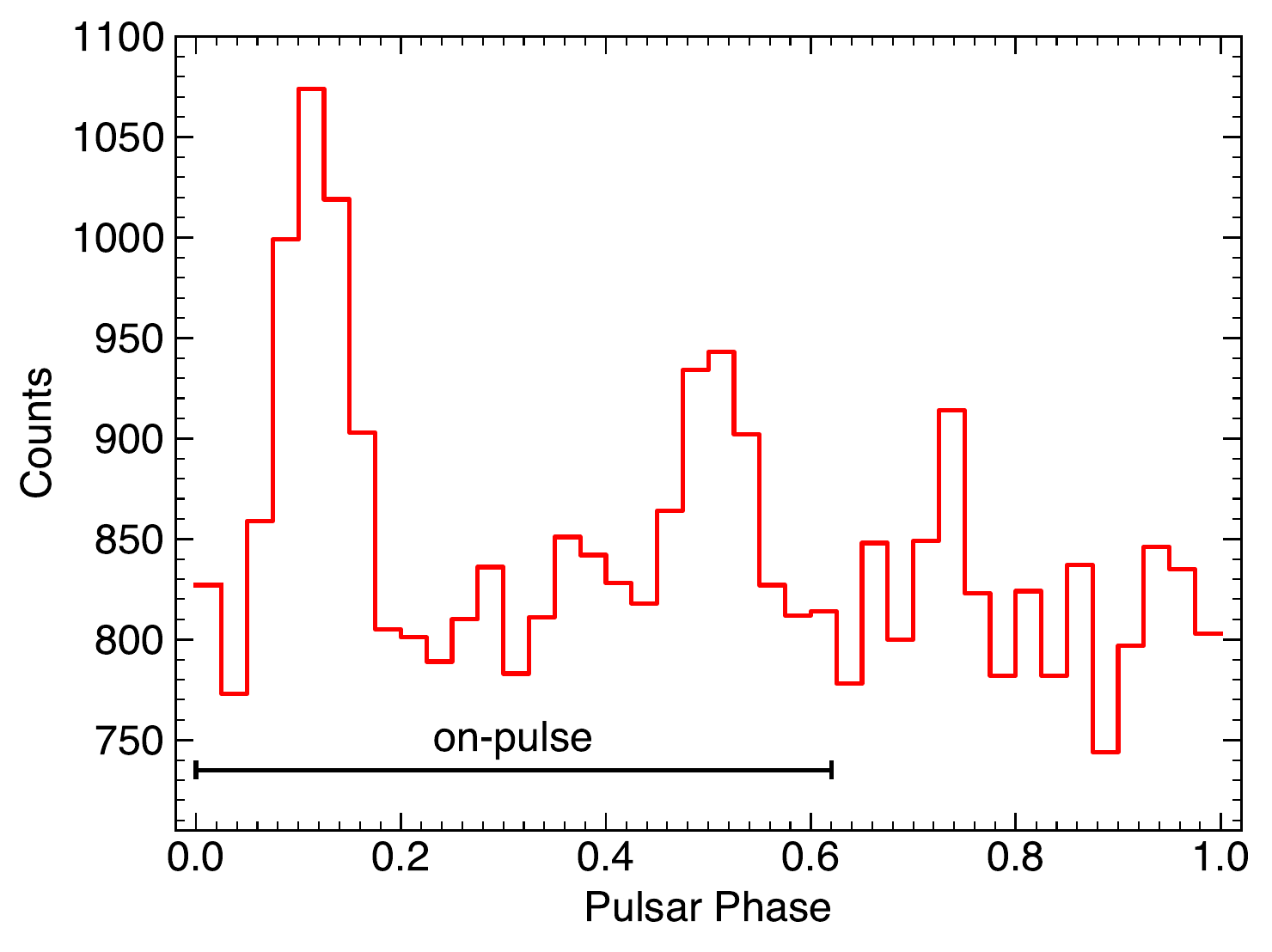}
\caption{Folded pulse profile of the Crab measured with PolarLight in the energy band of 3.0--4.5 keV. The on-pulse phase interval is indicated by the horizontal bar.
\label{fig:psr}}
\end{figure*}

%\clearpage

\begin{figure*}
\centering 
\includegraphics[width=0.5\columnwidth]{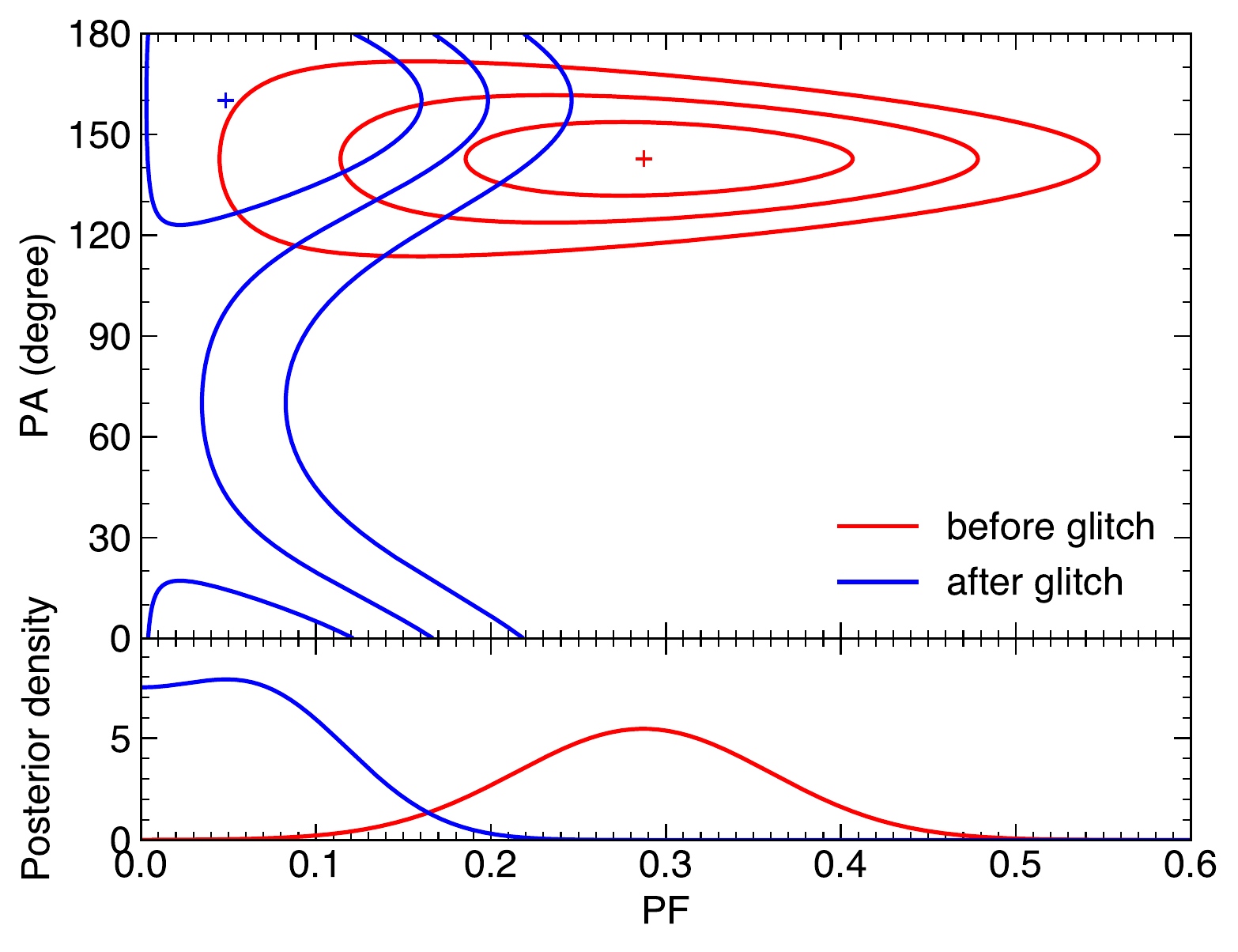}
\caption{Top: posterior distributions of PF and PA with data before the glitch (red) or data within 100~days after the glitch (blue).  Bottom: posterior distribution of PF (marginalised over PA).  Each measurement is not consistent with the other at a 3$\sigma$ level, and this conclusion is valid if one chooses any date from 30~days to 100~days after the glitch.
\label{fig:two_cont}}
\end{figure*}

%\clearpage

\begin{figure*}
\centering
\includegraphics[width=0.9\columnwidth]{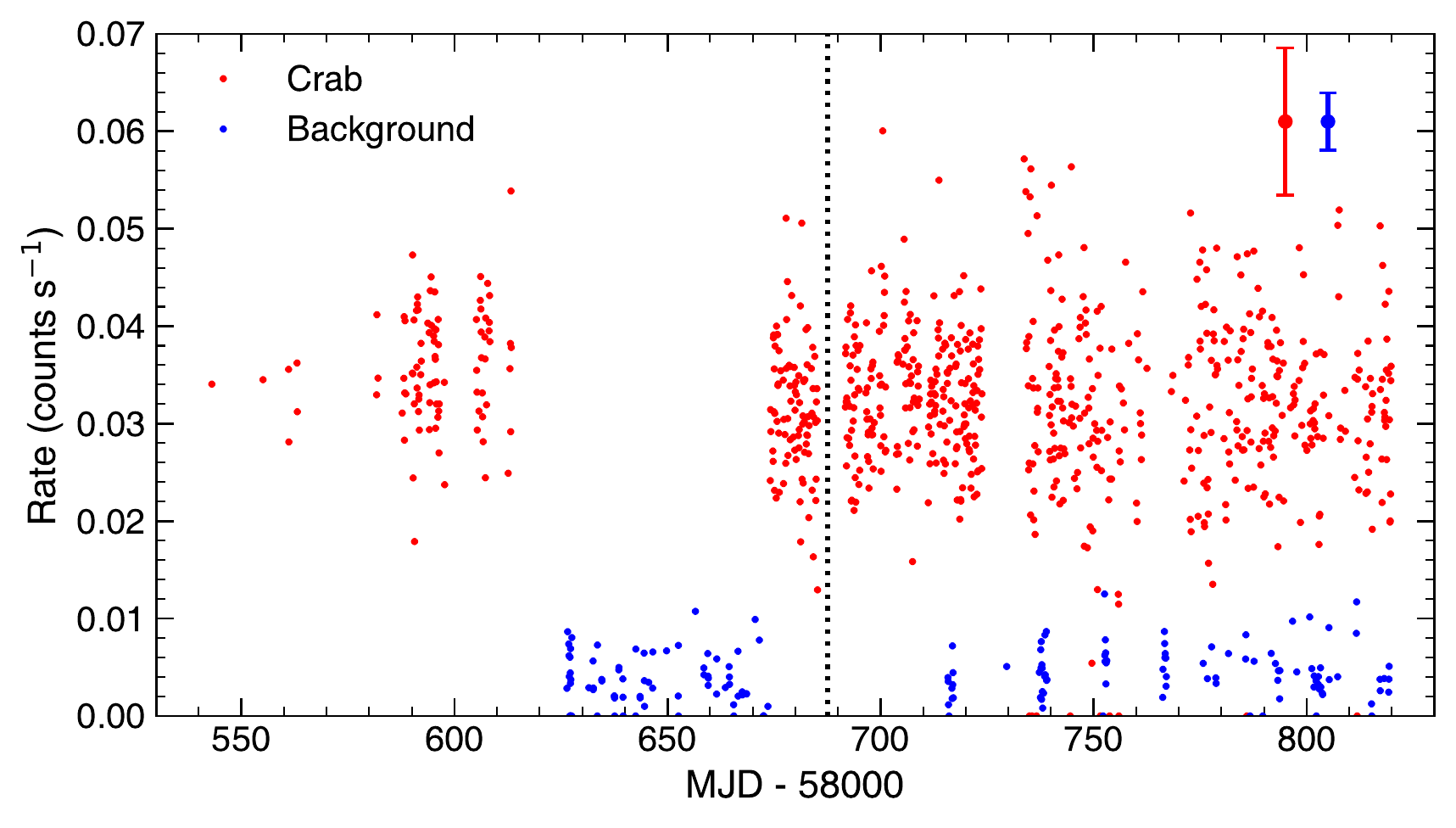}
\caption{3.0--4.5 keV lightcurves measured with PolarLight when observing the Crab and background regions. The bars show typical errors. Each point is the count rate averaged in a continuous exposure, which varies and has a typical duration of 15 minutes.  The gap in the Crab data from MJD 58620 to 58670 (early May to early July, 2019) is due to a small angular separation to the Sun, which precludes observations in this period.
\label{fig:lc}}
\end{figure*}
%%%%%%%%%%%%%%%%%%%%%%%%%%%%%%%%%%%%

\end{document}